\documentclass[11pt]{article}
\usepackage{graphicx}
\usepackage{newpasp}

\begin{document}


\title{Disks around Young Brown Dwarfs}

\author{Michael C. Liu\altaffilmark{1}}
\affil{Institute for Astronomy, 2680 Woodlawn Drive, University of
       Hawai\`{}i, Honolulu, HI 96822}
\altaffiltext{1}{Beatrice Watson Parrent Fellow. Email: mliu@ifa.hawaii.edu}


\begin{abstract}
We present some results from a systematic survey for disks around
spectroscopically identified young brown dwarfs and very low mass stars.
We find that $\approx$75\% of our sample show intrinsic IR excesses,
indicative of circum(sub)stellar disks.  The observed excesses are
well-correlated with H$\alpha$ emission, consistent with a common disk
accretion origin.  Because the excesses are modest, conventional
analyses using only IR colors would have missed most of the sources with
disks.  In the same star-forming regions, we find that disks around
brown dwarfs and T~Tauri stars are contemporaneous; assuming coevality,
this demonstrates that substellar disks are at least as long-lived as
stellar disks.  Altogether, the frequency and properties of
circumstellar disks are similar from the stellar regime down to the
substellar and planetary-mass regime. This offers compelling evidence of
a common origin for most stars and brown dwarfs.
\end{abstract}

\section{Introduction}

While the number of known brown dwarfs is growing rapidly, the origin of
these objects is an unanswered question.  One insight into the formation
mechanism(s) for brown dwarfs is whether young substellar objects
possess circumstellar disks.  There is abundant observational evidence
and theoretical expectation for accretion disks around young solar-type
stars.  Thus, the presence of disks around young brown dwarfs would be
naturally accommodated in ``star-like'' formation scenarios.  On the
other hand, scenarios involving dynamical interactions (e.g., collisions
and/or ejections) are likely to be hostile to circumstellar disks.

Evidence for circumstellar disks around individual young (few~Myr) brown
dwarfs has recently been found, from H$\alpha$ emission, near-IR
excesses, and mid-IR detections.  However, it is difficult to determine
the {\em frequency} of disks around brown dwarfs from studies to date
due to a combination of small number statistics, sample selection
inhomogeneity, and, most importantly, choice of wavelength.  A priori,
brown dwarf disks are expected to be harder to detect than disks around
stars because of lower contrast.  Substellar objects are less luminous
and have shallower gravitational potentials; hence, the inner regions of
their disks are likely to be cooler and thus could have negligible
excesses in the $JHK$ (1.1$-$2.4~\micron) bands, which have been used by
most previous studies.

\section{Observations}

We have recently completed a large $L^\prime$-band (3.8~\micron) survey
to study the frequency and properties of disks around young brown dwarfs
and very low mass stars (Liu, Najita, \& Tokunaga 2002).  Disks can be
readily identified by excess $L^\prime$-band emission, which arises from
warm material within a few stellar radii ($<$0.1~AU).  In addition, our
survey is sensitive enough to detect young brown dwarf photospheres, and
hence the absence of a disk can be discerned.  Our sample comprises
nearly all the published sources in Taurus and IC~348 which have been {\em
spectroscopically classified} to be very cool, with spectral types from
M6 to M9.5, corresponding to masses of $\sim$15 to $\sim$100~$M_{\rm
Jup}$ based on current models.  By focusing on targets with spectral
types, we are more sensitive to small IR excesses since we can determine
the intrinsic photospheric colors.  Our selection criterion also ensures
that we are targeting very low mass members, lying near or below the
stellar/substellar mass boundary (c.f., Muench et~al.\ 2001).

\begin{figure}[t]
\hspace{0.5cm}
\includegraphics[angle=90, width=4in]{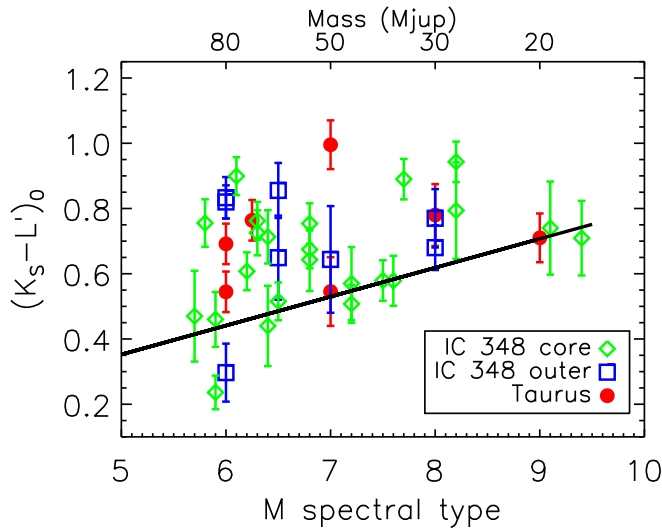}
\vskip -3ex
\caption{\small Dereddened $K_S\!-\!L^{\prime}$ colors as a function of
spectral type for our sample.  Typical errors in the spectral type are
$\approx$0.5 subclasses.  The heavy line represents the colors of field
M~dwarfs.  Most of the objects show intrinsic IR emission in excess of
that expected from their bare photospheres.  Approximate mass estimates
are listed at the top.}
\end{figure}

Figure~1 shows the dereddened $K_S\!-\!L^\prime$ (2.0$-$4.1~\micron)
colors of our sample compared to field (old, diskless) M~dwarfs.  Most
objects have colors which are redder than those expected from their bare
photospheres, i.e., they possess IR excesses. Also, the lower envelope
of the observed color distribution agrees well with the heavy line,
indicating that field M~dwarfs provide a legitimate comparison.
The disk frequency appears to be independent of mass; however, some
objects, including the very coolest (lowest mass) ones, lack IR
excesses.  We find an excess frequency of $\approx$75\% for the sample
as a whole --- disks around young brown dwarfs and VLM stars appear to
be very common.  The disk fraction in our sample is comparable to the
disk fraction for T~Tauri stars in the same star-forming regions (e.g.,
Strom et~al.\ 1989, Haisch et~al. 2001).


\section{IR Excesses of Young Brown Dwarfs}

{\em $\bullet$ H$\alpha$ Emission:} For T~Tauri stars, both the optical
line emission and the IR excesses are believed to originate from
disk accretion.  The IR excesses come from warm dust grains in the disk,
while H$\alpha$ emission arises from accretion of disk material onto the
central star.
Therefore, the H$\alpha$ emission and IR excesses should be correlated,
and indeed such a correlation is seen among T~Tauri stars (e.g., Kenyon
\& Hartmann 1995). Figure~2 shows a comparison for our sample, using
H$\alpha$ data from the literature.  A 3$\sigma$ correlation is observed
between the intrinsic $K_S\!-\!L^\prime$ excess and H$\alpha$ emission,
based on the Spearman rank correlation coefficient. This level of
correlation is comparable to that observed for T~Tauri stars and
provides strong circumstantial evidence for accretion disks around young
brown dwarfs.  Furthermore, the mere existence of accreting brown dwarfs
at ages of a few Myr argues for mass-dependent accretion rates, since
brown dwarfs with typical T~Tauri star accretion rates would not remain
substellar.

\begin{figure}[h]
\hspace{0.5cm}
\includegraphics[angle=90, width=4in]{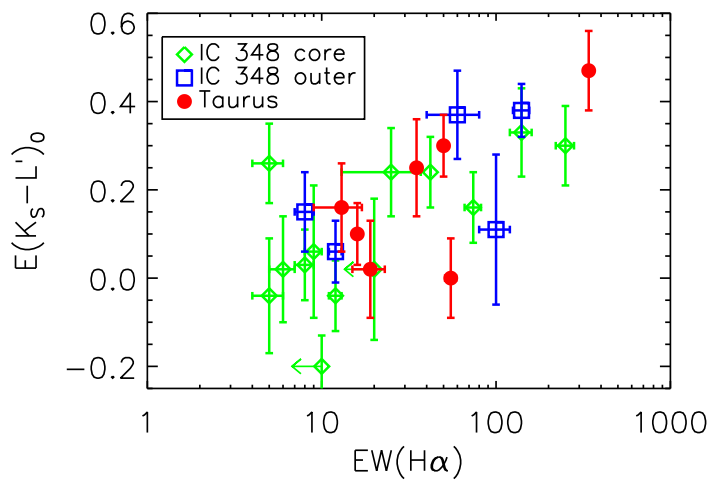}
\vskip -3ex
\caption{\small Relation between intrinsic $K_S\!-\!L^\prime$ excess and
H$\alpha$ equivalent width in Angstroms for most of our sample.  The two
quantities are well-correlated, supporting the idea that the optical and
near-IR emission both originate from the same phenomenon, namely
circumstellar accretion disks.}
\end{figure}


\noindent {\em $\bullet$ Finding Stellar and Substellar Disks:} The most
common method for measuring the disk fraction of T~Tauri stars uses
color-color diagrams based on $JHK$, or preferably $JHKL$, colors.
Objects with disks are identified as those having colors distinct from
reddened dwarf and giant stars.
This method is appealing since only photometry is required.  Figure~3
illustrates this method as applied to our sample.  From this diagram,
one would identify only $\approx$1/3 of the objects as having IR
excesses. However, our analysis which incorporates the objects' spectral
types shows that in fact the majority of sources with disks are missed
in the color-color diagram because their excesses are modest.

The reason why the conventional analysis works poorly is because of
decreased contrast between brown dwarfs and their disks, as compared to
the case of T~Tauri stars.  Because brown dwarfs are less luminous than
the higher mass T~Tauri stars, passive disks around brown dwarfs will be
cooler and less luminous. Therefore, the corresponding IR excesses will
be smaller.  This physical intuition is verified with simple disk models
(see Liu et~al.\ 2002 for details).


\begin{figure}
\hbox{
\vbox{\hsize=3.5truein
\vskip -0.4in
\hskip -0.5in
\includegraphics[angle=90, width=4.3in]{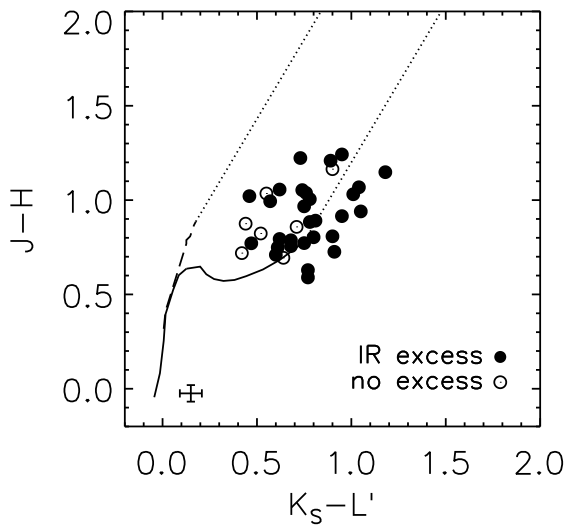}
}
\raise 0.45in
\hbox{\hskip -0.7in
\vbox{\hsize=2.6truein 
\caption{\small Using a conventional color-color analysis, only a small
fraction of the sample would be identified as having IR excesses.
However, when incorporating the spectral types, many more are found to
have excesses ($\bullet$).  Most of these are missed in the color-color
diagram because their IR excesses are modest.  The various lines
represent the loci of reddened giant (G0 to M5) and dwarf (A0 to M9.5)
stars.  Median errors are plotted in the lower right.}}}}
\end{figure}

\medskip
\noindent {\em $\bullet$ Implications:} We find that (1)~disks around
brown dwarfs are common, and (2)~brown dwarf disks are contemporaneous
with disks around T~Tauri stars in the same star-forming regions.  The
latter also shows that brown dwarf disks are at least as long-lived as
disks around stars, assuming that the stars and brown dwarfs are roughly
coeval.  These observations are naturally accommodated in a picture
where brown dwarfs form in a similar manner as stars --- our results
offer prima facie evidence for a common origin for objects from the
stellar regime down to the substellar and planetary-mass regime.

Alternative formation scenarios, such as disk-disk collisions and/or
premature ejection, involve dynamical interactions in creating brown
dwarfs.  While specific predictions are lacking due to the stochastic
nature of these scenarios, brown dwarfs formed by these mechanisms are
generally expected to have smaller and less massive disks, and
consequently shorter disk lifetimes, compared to brown dwarfs formed in
isolation.  This expectation conflicts with our finding that brown dwarf
disks are at least as long-lived as disks around young stars.


\acknowledgments I thank my collaborators, Joan Najita and
Alan Tokunaga, for valuable input.  I also thank the IAU 211 SOC and LOC
for organizing a very enjoyable conference.



\end{document}